\documentclass{optica-article}

\journal{opticajournal} 

\articletype{Research Article}

\usepackage{lineno}

\begin{document}

\title{Transfer-Function Approach to Substrate-Enhanced Diffraction Tomography}

\author{Tongyu Li,\authormark{1} Yi Shen,\authormark{1} Dashan Dong,\authormark{1} Danchen Jia,\authormark{1} Jianpeng Ao,\authormark{1} Ji-Xin Cheng,\authormark{1, 2} and Lei Tian\authormark{1,2,*}}

\address{\authormark{1}Department of Electrical and Computer Engineering, Boston University, Boston, Massachusetts 02215, USA\\
\authormark{2}Department of Biomedical Engineering, Boston University, Boston, Massachusetts 02215, USA}

\email{\authormark{*}leitian@bu.edu} 


\begin{abstract*}
Forward and backward scattering provide complementary volumetric and interfacial information, yet conventional three-dimensional (3D) imaging typically accesses only one. We present a substrate-enhanced diffraction tomography approach that simultaneously recovers both channels under multi-angle epi-illumination. This geometry captures one forward- and two backward-scattering bands in axially symmetric Fourier regions, where their complementary coverage enables phase–absorption separation in a non-Hermitian spectrum.  Explicit 3D transfer functions are derived for both channels, and an axial Kramers–Kronig relation is established to incorporate substrate-induced boundary conditions in a unified framework. Our results establish a label-free, high-resolution 3D imaging modality that surpasses the limits of existing methods.
\end{abstract*}

\section{Introduction}

3D imaging through a finite numerical aperture (NA) objective lens is fundamentally constrained by anisotropic spatial bandwidth, yielding high lateral but poor axial resolution. 
Forward scattering (FS) predominantly encodes volumetric information but suffers from the missing-cone problem, where high-axial-frequency components are inaccessible~\cite{park2018quantitative}. 
Backward scattering (BS), in contrast, carries high-frequency interfacial features but does not provide quantitative volumetric information~\cite{huang1991optical, uttam2015fourier}. 
These channels are inherently complementary, yet are conventionally probed in separate transmission or reflection geometries~\cite{jin2017tomographic,foucault2019versatile}. Extending the accessible 3D scattering spectrum is thus essential for accurate characterization, but existing solutions, such as dual opposing objective lenses~\cite{hell1994confocal}, mechanical sample rotation~\cite{kus2014tomographic, simon2017tomographic, moser2025optical} or engineered reflectors~\cite{zhou2022computational}, remain impractical.

Among these strategies, a reflective substrate (e.g., a mirror) offers an efficient means of folding counter-propagating scattering fields into a common pupil, effectively emulating opposing objectives within a single-objective geometry~\cite{bailey1993enhancement}. Under appropriate NA conditions, such configurations can, in principle, approach 4Pi tomographic bandwidth~\cite{mudry2010mirror}.
Building on this concept, we advance a substrate-enhanced diffraction tomography technique that simultaneously captures FS and BS from \emph{intensity-only} measurements~\cite{tian20153d,chowdhury2019high} using a multi-angle epi-illumination configuration~\cite{maire2009experimental, mudry2010mirror, mudry2010isotropic, li2025reflection}. 
This configuration further eliminates the need for a dedicated interferometric path and effectively forms a dual-view acquisition~\cite{chacon2025dual}, expanding the accessible 3D spatial frequency bandwidth threefold [Fig.~\ref{fig:overview}(a)]. 
In this arrangement, both channels interfere with the reflected incidence, providing a natural self-reference that enables quantitative phase reconstruction from intensity-only measurements.
This substrate-enhanced acquisition strategy provides access to axially symmetric Fourier bands---one corresponding to FS and two to BS---whose complementary coverage enables phase–absorption separation in a non-Hermitian spatial frequency spectrum. In contrast, conventional reflection geometries capture only a single BS band, which is insufficient for such separation~\cite{uttam2015fourier, kang2023mapping}.

A central challenge in recovery is that FS and BS contributions from different depths overlap in the recorded 2D intensity patterns. To disentangle them, we employ diffraction tomography with angle-diverse measurements [Fig.~\ref{fig:overview}(b–c)], which reconstructs the 3D scattering potential by solving an inverse scattering problem. 
Our previous work employed the modified Born series (MBS)~\cite{osnabrugge2016convergent} for this task~\cite{li2025reflection}, and while effective, such rigorous models, including the discrete dipole approximation~\cite{mudry2010mirror, zhang2013full}, require computationally intensive iterative reconstructions that obscure physical insight.
To address this, we derive explicit 3D transfer functions under the first-Born approximation, which establish a linear relationship between the measured intensity and the underlying scattering potential.
These transfer functions confine the reconstruction to the physical passbands, substantially reducing computational cost.
Moreover, the Kramers–Kronig (KK) relation, originally established in the temporal domain, has recently been extended to the lateral spatial domain for efficient phase retrieval~\cite{baek2019kramers, baek2021intensity}.
Here, we further exploit the spatial causality of scattering in the presence of a reflective substrate to establish a new axial KK relation, which constrains the inversion of the half-space scattering problem.
Together, this transfer-function framework provides a clearer and more tractable alternative to iterative MBS model, enabling efficient separation and recovery of 3D FS and BS information, as validated in simulation and experiment.

\section{Theory and method}
\begin{figure}[t]
\centering
\includegraphics[width=8.5cm]{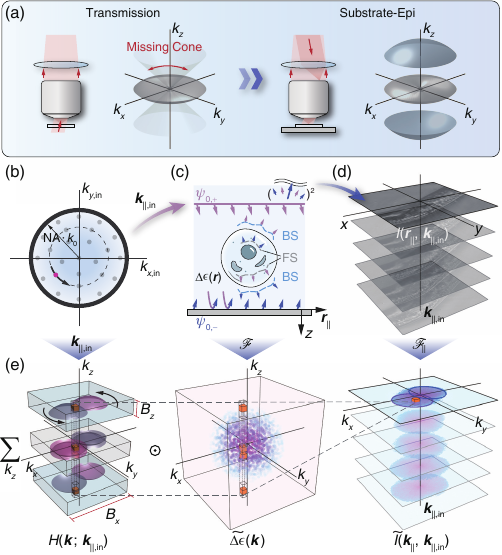}
\caption{\label{fig:overview} 
(a)~Passbands in transmission and substrate-enhanced epi configurations; the latter simultaneously records FS and BS, expanding the accessible spectrum threefold.
(b)~Angle-diverse illumination with varying ${\mathbf{k}}_{\parallel,\mathrm{in}}$ enables retrieval of 3D FS and BS.
(c)~Substrate-enhanced epi-illumination schematic, the incident field and its substrate reflection act as dual illuminations generating FS and BS.
(d)~Representative intensity patterns measured under different illumination angles.
(e)~Linear scattering model expressed via the 3D transfer function.}
\end{figure}

Consider a 3D weakly-scattering object immersed in a medium permittivity $\epsilon_{\mathrm{m}}$ above a reflective substrate, characterized by a permittivity perturbation $\Delta\epsilon(\mathbf{r}) = \epsilon(\mathbf{r}) - \epsilon_{\mathrm{m}}$. 
The substrate surface is defined as the $z = 0$ plane, with its optical response described by a Fresnel coefficient $R(\mathbf{k}_{\parallel})$. 
An obliquely incident plane wave $\psi_{0, +} = e^{i(\mathbf{k}_{\parallel, \mathrm{in}}\cdot \mathbf{r}_{\parallel}+k_{z, \mathrm{in}} z)}$ with unit amplitude, propagating with $k_{z, \mathrm{in}}$ along the positive $z$-direction, illuminates the object and substrate, as illustrated in Fig.~\ref{fig:overview}(c).

Under the single-scattering approximation, the reflective substrate first reflects the incidence back toward the object, generating a secondary incidence, $\psi_{0, -} = R(\mathbf{k}_{\parallel, \mathrm{in}})e^{i(\mathbf{k}_{\parallel, \mathrm{in}}\cdot \mathbf{r}_{\parallel}-k_{z, \mathrm{in}} z)}$, forming a standing-wave illumination.
Each incidence scatters at every depth, producing FS along the incidence (gray dashed line) and BS along the opposite direction (light blue dashed line). 
For $\psi_{0,+}$ (purple arrows), FS ($\psi_{\mathrm{FS},+}$) propagates along $+z$ and BS ($\psi_{\mathrm{BS},+}$) along $-z$; 
For $\psi_{0,-}$ (blue arrows), FS ($\psi_{\mathrm{FS},-}$) propagates along $-z$ while BS ($\psi_{\mathrm{BS},-}$) propagates along $+z$. 
Scattered fields propagating along $-z$ are directly collected, while $+z$-propagating fields first reach the substrate, reflect at $z=0$ to reverse the direction, and then enter the objective. 
In total, the model includes four scattered field components: $\psi_{\mathrm{FS},-}$ and $\psi_{\mathrm{BS},+}$, corresponding to the FS and BS of the real image, and $\psi_{\mathrm{FS},+}$ and $\psi_{\mathrm{BS},-}$, corresponding to the FS and BS of the virtual image generated by substrate reflection.

Without loss of generality~\cite{li2025reflection}, the objective's focal plane is set at $z=0$. The microscope effectively applies a Fourier filter to the field through its pupil function $P(\mathbf{k}_{\parallel})$. 
The secondary incidence $\psi_{0,-}$ interferes with all four scattered field components to form the measured intensity pattern $I_{\mathbf{k}_{\parallel, \mathrm{in}}}(\mathbf{r}_{\parallel})$.
Under the weak-scattering condition, the cross terms provide the main interference contrast, establishing a linear relation between the permittivity contrast and the intensity spectrum~\cite{ling2018high},
\begin{equation}\label{eq:slice-wiseTF}
    \tilde{I}_{\mathbf{k}_{\parallel, \mathrm{in}}}({\mathbf{k}_{\parallel}}) \approx
    \tilde{I}_{0, \mathbf{k}_{\parallel, \mathrm{in}}} +\int_{-\infty}^0 \mathbf{H}_{ \parallel,\mathbf{k}_{\parallel, \mathrm{in}}}({\mathbf{k}_{\parallel}},z)\widetilde{\Delta \boldsymbol{\epsilon} }_{\parallel}({\mathbf{k}_{\parallel}},z)\cdot\mathrm{d}z,
\end{equation}
where $\mathbf{H}_{\parallel} = [H_{\mathrm{\parallel,Re}} ,H_{\parallel,\mathrm{Im}} ]$ denotes the slice-wise transfer function, $\widetilde{\Delta \boldsymbol{\epsilon} }_{\parallel} = [\widetilde{\Delta \epsilon }_{\parallel,\mathrm{Re}}, \widetilde{\Delta \epsilon }_{\parallel,\mathrm{Im}}]^\top$ is the 2D spectrum of each slice, and $\tilde{I}_{0, \mathbf{k}_{\parallel, \mathrm{in}}}$  is the direct current component, which can be removed by background subtraction~\cite{ling2018high}.

Unlike transmission geometry, where scattering occurs throughout the space unless previously bounded, the epi-configuration restricts the object to the upper half-space ($z<0$), leaving the region $z>0$ free of scatterers. This asymmetry prevents a direct definition of a full 3D transfer function. However, the substrate-imposed cutoff naturally enforces a spatial causality constraint, where the scattering potential is confined to one half-space in real space, and its Fourier dual is analyticity in $k_z$. This analyticity ensures that all scattered fields originate from the physical half-space and leads to an axial KK relation, linking the real and imaginary parts of $\widetilde{\Delta \epsilon}$ through a Hilbert transform $\mathscr{H}_z$ along $k_z$,
\begin{equation}
\label{KK}
    \mathrm{Im}[\widetilde{\Delta \epsilon }({\mathbf{k}})] =
    \frac{1}{\pi} \mathrm{p.v.}\int_{-\infty}^\infty \frac{\mathrm{Re}[\widetilde{\Delta \epsilon }({\mathbf{k}}_{\parallel}, k'_z)]}{k_z - k'_z} \,\mathrm{d}k'_z,
\end{equation}
where p.v. denotes the Cauchy principal value, in direct analogy to the temporal KK relations that connect a material’s dispersive permittivity to causality.

\begin{figure}[t]
\centering
\includegraphics[width=8.5cm]{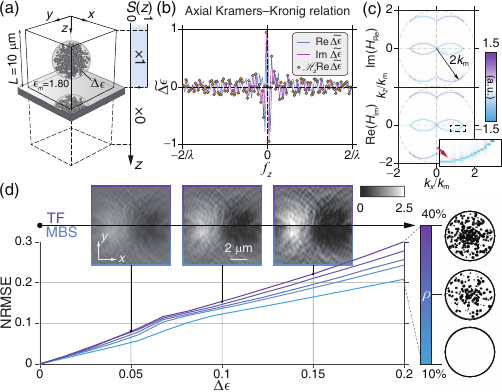}
\caption{\label{fig:TFvalidationz} 
(a)~Synthetic object above a mirror substrate. The mirror acts as a step function, enforcing causality in the scattering spectra.
(b)~Real and imaginary parts of the normalized object spectrum at $\mathbf{k}_{\parallel}=(0,0)$; yellow points show $\mathscr{H}_z[\mathrm{Re}(\widetilde{\Delta\epsilon})]$ matching $\mathrm{Im}(\widetilde{\Delta\epsilon})$, demonstrating the axial KK relation.
(c)~$k_x$–$k_z$ profile of the 3D transfer function for a 0.95 NA under NA-matched incidence, $k_m = \epsilon^{1/2}_mk_0$.
(d)~NRMSE between the transfer function and MBS models as a function of $\Delta \epsilon$ and volume fraction; insets show representative computed patterns at increasing $\Delta \epsilon$.
}
\end{figure}

We make this property explicit by expressing $\Delta \epsilon({\mathbf{r}}) \to S(z)\Delta \epsilon({\mathbf{r}})$, where $S(z)$ is a step function enforcing analyticity [Fig.~\ref{fig:TFvalidationz}(a) and (b)]. This formalizes the causal nature of the scattering process and permits the integral in Eq.~\eqref{eq:slice-wiseTF} to be extended over the entire space $(-\infty,\infty)$ without changing the physical content. Using the Parseval–Plancherel identity, scattering from an object above a reflective substrate can then be expressed linearly through a 3D transfer function,
\begin{equation}\label{eq:3DTF}
    \tilde{I}_{\mathbf{k}_{\parallel, \mathrm{in}}}({\mathbf{k}_{\parallel}}) \approx
    \tilde{I}_{0, \mathbf{k}_{\parallel, \mathrm{in}}} +\int_{-\infty}^\infty \mathbf{H}_{\mathbf{k}_{\parallel, \mathrm{in}}}({\mathbf{k}})\widetilde{\Delta \boldsymbol{\epsilon} }({\mathbf{k}})\cdot\mathrm{d}k_z,
\end{equation}
where
\begin{subequations}
    \begin{align}
    \label{eq:TF4Phase}
        H_{\mathrm{Re}, \mathbf{k}_{\parallel, \mathrm{in}}}(\mathbf{k}) &= \frac{ik_0^2}{2}\left[ H_{0, \mathbf{k}_{\parallel, \mathrm{in}}}(\mathbf{k}) - H^*_{0, \mathbf{k}_{\parallel, \mathrm{in}}}(-\mathbf{k}) \right],\\
    \label{eq:TF4Absorb}
        H_{\mathrm{Im}, \mathbf{k}_{\parallel, \mathrm{in}}}(\mathbf{k}) &= -\frac{k_0^2}{2}\left[ H_{0, \mathbf{k}_{\parallel, \mathrm{in}}}(\mathbf{k}) + H^*_{0, \mathbf{k}_{\parallel, \mathrm{in}}}(-\mathbf{k}) \right].
    \end{align}
\end{subequations}
Here,
$H_{0, \mathbf{k}_{\parallel, \mathrm{in}}}(\mathbf{k})$ contains the essential contributions,  consisting of two FS and two BS components,
\begin{equation}
     H_{0, \mathbf{k}_{\parallel, \mathrm{in}}}(\mathbf{k}) = \frac{R^*({\mathbf{k}_{\parallel, \mathrm{in}}})P^*({\mathbf{k}_{\parallel, \mathrm{in}}})P(\mathbf{k}_{\parallel} + \mathbf{k}_{\parallel, \mathrm{in}} ) }{k_{\bot}({\mathbf{k}_{\parallel}}+{\mathbf{k}_{\parallel, \mathrm{in}}})} (H_{\mathrm{FS}} + H_{\mathrm{BS}} ),
\end{equation}
with the FS and BS components
\begin{subequations}
\begin{align}
\label{HFS}
H_{\mathrm{FS}} &= 
R(\mathbf{k}_{\parallel} + \mathbf{k}_{\parallel, \mathrm{in}} )\delta_{\mathrm{FS},+}
+ R(\mathbf{k}_{\parallel, \mathrm{in}})\delta_{\mathrm{FS},-}, \\
\label{HBS}
H_{\mathrm{BS}} &=
\delta_{\mathrm{BS},+}
+ R(\mathbf{k}_{\parallel, \mathrm{in}})
  R(\mathbf{k}_{\parallel} + \mathbf{k}_{\parallel, \mathrm{in}})
  \delta_{\mathrm{BS}, -},
\end{align}
\end{subequations}
where $k_{\bot}({\mathbf{k}_{\parallel}}+{\mathbf{k}_{\parallel, \mathrm{in}}})\equiv [ \epsilon_{\mathrm{m}} k_0^2- ({\mathbf{k}_{\parallel}}+{\mathbf{k}_{\parallel, \mathrm{in}}})^2]^{1/2}$, $\delta_{\mathrm{FS},+/-} \equiv \delta[k_z \mp k_{\bot}({\mathbf{k}_{\parallel}}+{\mathbf{k}_{\parallel, \mathrm{in}}}) \pm k_{z, \mathrm{in}} ]$ and $\delta_{\mathrm{BS},+/-} \equiv \delta[k_z \pm k_{\bot}({\mathbf{k}_{\parallel}}+{\mathbf{k}_{\parallel, \mathrm{in}}}) \pm k_{z, \mathrm{in}} ]$. Details in Supplement 1.

Equations~\eqref{eq:TF4Phase} and \eqref{eq:TF4Absorb} show that $\widetilde{\Delta \epsilon }_{\mathrm{Re}}({\mathbf{k}})$ and $\widetilde{\Delta \epsilon }_{\mathrm{Im}}({\mathbf{k}})$ are encoded asymmetrically in Fourier domain: 
$H_{\mathrm{Re}, \mathbf{k}_{\parallel, \mathrm{in}}}$ is imaginary and anti-symmetric, while $H_{\mathrm{Im}, \mathbf{k}_{\parallel, \mathrm{in}}}$ is real and symmetric.
The substrate effectively doubles the accessible Fourier domain along $k_z$ by folding $+z$-propagating FS and BS components into the pupil, effectively forming a dual-view acquisition.
This yields four $\delta_{\mathrm{FS}/\mathrm{BS},+/-}$ terms that define Ewald spherical caps centered at $\mathbf{k}_{\parallel,\mathrm{in}}$, truncated by the NA, and separated symmetrically along $k_z$, forming Fourier pairs that enable separation of phase and absorption in both FS and BS from any non-Hermitian spectrum. [Equations~\eqref{HFS} and \eqref{HBS}; Fig.~\ref{fig:overview}(e)].

This asymmetry and separation imply that phase and absorption, as well as FS and BS contributions, are inherently decoupled and thus separable during retrieval. As the illumination angle varies, FS and BS caps sweep through Fourier domain and fill their respective supports [Fig.~\ref{fig:overview}(a)], yielding three isolated bands with lateral bandwidths $B_x = B_y = 4\mathrm{NA}\cdot k_0$ and axial bandwidth $B_z = 2\epsilon^{1/2}_\mathrm{m}k_0(1 - \sqrt{1 - \mathrm{NA}^2/\epsilon_{\mathrm{m}}})$~\cite{sarmis2010high}. 
Two BS bands (green) and the FS band (gray) are highlighted in Fig.~\ref{fig:overview}(e).

The inverse scattering problem seeks to recover the unknown $\Delta \epsilon ({\mathbf{r}})$ of the object from a series of angle-diverse measurements. The forward model, defined in Eq.~\eqref{eq:3DTF}, is linear and decouples across in-plane spatial frequencies ${\mathbf{k}_{\parallel}}$. In the discrete formulation, we assume sampling in all three dimensions meets the Nyquist criterion imposed by the respective bandwidths.

At a given ${\mathbf{k}_{\parallel}}$, the key challenge is to disentangle FS and BS components, which are superimposed in the forward model. 
To address this, we collect the full stack of $M$ angle-diverse measurements, leading to a discrete linear system [Fig.~\ref{fig:overview}(e)],
\begin{equation}
    \mathbf{A}({\mathbf{k}_{\parallel}})
    \widetilde{\Delta \boldsymbol{\epsilon} }({\mathbf{k}_{\parallel}})
    = \tilde{\mathbf{I}}({\mathbf{k}_{\parallel}}),
\end{equation}
where $\tilde{\mathbf{I}}({\mathbf{k}_{\parallel}})\in \mathbb{R}^{M\times 1}$ is the column vector of measured intensity spectra at ${\mathbf{k}_{\parallel}}$,
$\widetilde{\Delta \boldsymbol{\epsilon} }({\mathbf{k}_{\parallel}}) \in \mathbb{C} ^ {2N \times 1}$, given by $[\widetilde{\Delta \boldsymbol{\epsilon} }^{\top}_{\mathrm{Re}}({\mathbf{k}_{\parallel}}) , \widetilde{\Delta \boldsymbol{\epsilon} }^{\top}_{\mathrm{Im}}({\mathbf{k}_{\parallel}}) ]^{\top}$, contains the unknown Fourier components of the permittivity contrast within the passband (for $N$ axial frequency components), and forward operator $\mathbf{A}\in\mathbb{C}^{M\times 2N}$ mapping $\widetilde{\Delta \boldsymbol{\epsilon}}$ to the measurement spectra across all illuminations,
\begin{equation}\label{eq:ForwardMatrix}
   \mathbf{A} \equiv \left(\mathbf{H}_{\textrm{Re}},\mathbf{H}_{\textrm{Im}}\right)
    \begin{pmatrix}
       (1+i\mathscr{H}_z)/2&0\\0 & (1+i\mathscr{H}_z)/2
    \end{pmatrix},
\end{equation}
where the axial KK relation in Eq.~\eqref{KK} is imposed by invoking the identity $(1+i\mathscr{H}_z)\widetilde{\Delta \boldsymbol{\epsilon} }({\mathbf{k}_{\parallel}})/2 = \widetilde{\Delta \boldsymbol{\epsilon} }({\mathbf{k}_{\parallel}})$.
In practice, we apply discrete Fourier transforms (DFTs) to the discretized slice-wise transfer functions [Eq.~\eqref{eq:slice-wiseTF}] to compute the 3D transfer function. This approach is numerically more stable than direct discretization of Equations~\eqref{eq:TF4Phase} and \eqref{eq:TF4Absorb}, as it naturally enforces the Nyquist bandlimit and properly handles cases where delta locations fall off the Fourier sampling grid. Example 3D transfer functions are shown in Fig.~\ref{fig:TFvalidationz}(c), and details are shown in Supplement 1.

To validate the forward model, we simulated scattering from a synthetic object placed above a mirror under oblique illumination [Fig.~\ref{fig:TFvalidationz}(a)]. A mirror is used as the substrate to maximize collection efficiency, with the Fresnel coefficient $R(\mathbf{k}_{\parallel})=-1$. The object consists of randomly distributed beads confined to a thin spherical shell (radius 3 $\mu$m, thickness 100 nm) with volume fraction $\rho=10$–40\% and permittivity contrast $\Delta \epsilon=0$–0.2 on a background of $\epsilon_{\mathrm{m}}=1.80$, illuminated by a 632 nm plane wave with $\mathbf{k}_{\parallel, \mathrm{in}}=(0.6k_0,0)$. The normalized root-mean-square error (NRMSE) between the computed patterns using the transfer function approach and the rigorous MBS model [Fig.~\ref{fig:TFvalidationz}(d)] remains below 0.3 across all tested parameters, confirming that Eq.~\eqref{eq:3DTF} accurately captures both forward- and backward-scattered fields. Representative patterns at $\Delta \epsilon=0.05,~0.1,$ and $0.15$ (inset) show that Eq.~\eqref{eq:3DTF} reproduces the interference features of the rigorous MBS model with consistent contrast variations as $\Delta \epsilon$ increases. The error grows nearly linearly with $\Delta \epsilon$, reflecting stronger multiple scattering beyond the single-scattering regime, yet remains quantitatively reliable for typical biomedical imaging conditions.

For recovering 3D FS and BS information, the linear forward model enables efficient non-iterative inversion, while the axial KK relation enforces spatial causality as a physical prior, confining the solution to the upper half-space.
For practical implementation, rather than discretizing the entire real space at the Nyquist limit as required by our previous MBS method~\cite{li2025reflection}, the axial separation of FS and BS components in the 3D Fourier space allows a band-limited reconstruction confined to the three Fourier-domain supports illustrated in Fig.~\ref{fig:overview}(e). 
This restriction reduces the dimensionality of the forward operator $\mathbf{A}$ to the axial passband, substantially lowering the memory cost of inversion. 
The inverse is computed using truncated singular value decomposition, with $\mathbf{A}^{+} \approx \mathbf{V}\mathbf{\Sigma}_{\alpha}^{-1}\mathbf{U}^\dagger$ constructed by discarding singular values below a tunable threshold $\alpha$, thereby suppressing low signal-to-noise (SNR) modes while preserving the dominant scattering components, providing regularization analogous to an $l_2$-type constraint.

\section{Numerical validation}

\subsection{Complementary FS–BS contrast and phase-absorption separation}

\begin{figure}[b]
\hspace{-2cm}
\includegraphics[width=17cm]{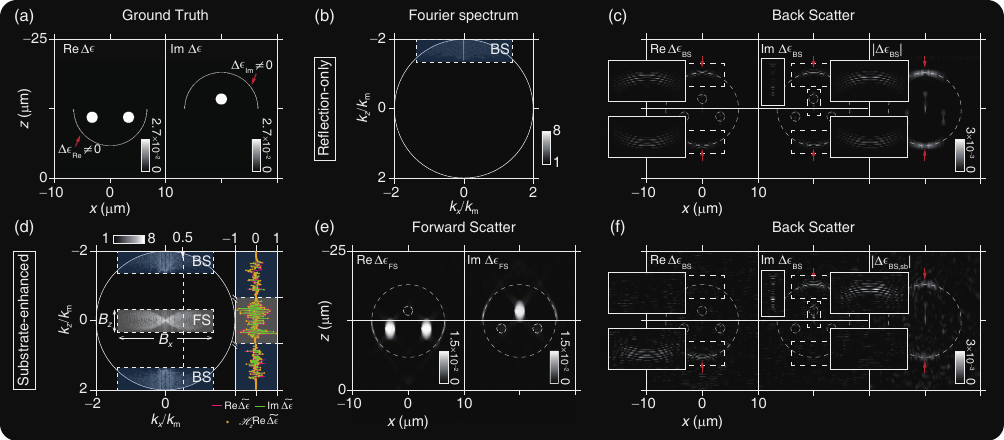}
\caption{\label{fig:SimulationRecon}
Simulation validation of the reconstruction framework.
(a)~Ground truth of the synthetic object, with the lower half purely phase and the upper half purely absorptive.
(b)~$k_x$–$k_z$ profile of the recovered $\widetilde{\Delta\epsilon}$ from reflection-only geometry.
(c)~$x$–$z$ cross-sections of the reconstructed $\Delta\epsilon_{\mathrm{BS}}$ from reflection-only geometry.
(d)~$k_x$–$k_z$ profile of the recovered $\widetilde{\Delta\epsilon}$ from substrate-enhanced geometry; right panel: cross-section at $k_x=0.5k_m$ showing $\mathrm{Re}\,\widetilde{\Delta\epsilon}$ and $\mathrm{Im}\,\widetilde{\Delta\epsilon}$ satisfy the axial KK relation.
(e)~$x$–$z$ cross-sections of the real and imaginary parts of $\Delta\epsilon_{\mathrm{FS}}$ obtained from the FS band.
(f)~$x$–$z$ cross-sections of the real and imaginary parts of $\Delta\epsilon_{\mathrm{BS}}$ and $|\Delta\epsilon_{\text{BS, sb}}|$ obtained from the combined BS bands and the single BS band, respectively.
$\Delta\epsilon_{\mathrm{FS}}$ reveals smooth volumetric features with blurred interfaces, while $\Delta\epsilon_{\mathrm{BS}}$ complements FS by recovering lateral interfaces, with their real and imaginary parts corresponding to the decoupled phase and absorption, respectively.
Figures are clipped to positive values for presentation.
}
\end{figure}

We first validate the method in simulation using an object with $|\Delta\epsilon|=2.7\times10^{-2}$, consisting of three beads (radius 1~$\mu$m) embedded in a thin spherical shell (radius 6.5~$\mu$m, thickness 100 nm) on a background of $\epsilon_{\mathrm{m}}=1.80$, as shown in Fig.~\ref{fig:SimulationRecon}(a). The beads are used to evaluate the spatial elongation, while the shell interfaces provide strong sensitivity to BS.
The top bead and upper half-shell are purely absorptive ($\Delta\epsilon_{\mathrm{Im}} \neq 0$, $\Delta\epsilon_{\mathrm{Re}} = 0$), whereas the bottom beads and lower half-shell are purely phase objects ($\Delta\epsilon_{\mathrm{Re}} \neq 0$, $\Delta\epsilon_{\mathrm{Im}} = 0$).
Angle-diverse intensity measurements are simulated using the rigorous MBS model with 150 illumination angles at 632 nm, sampled along a Fermat's spiral and spanning a maximum illumination angle of 0.95, with simulation details shown in Supplement 1.
The reconstruction results under substrate-enhanced geometry using the proposed method are shown in Fig.~\ref{fig:SimulationRecon}(d--f). For comparison, the reflection-only reconstruction from simulations without the substrate are shown in Fig.~\ref{fig:SimulationRecon}(b--c).

In the substrate-enhanced geometry, three recovered bands are assembled into the 3D Fourier spectrum after inversion. Figure~\ref{fig:SimulationRecon}(d) shows the $k_x$–$k_z$ cross-section of the logarithmic Fourier spectrum, while the $k_x=0.5k_m$ slice on the right confirms the axial KK relation of the recovered spectrum. 

Because the spectrum is non-Hermitian, regions symmetric about the axial frequency $k_z$ must be combined to decouple phase and absorption. For the FS band, centered at zero frequency with support inherently symmetric about $k_z=0$, this symmetry alone suffices for decoupling, corresponding to $\Delta\epsilon_{\text{FS}}$ in real space. By contrast, BS bands occupy two distinct high-frequency regions symmetric about the $k_z$ axis in substrate-enhanced geometry; both must be combined to achieve decoupling, with their real and imaginary parts ($\text{Re}\,\Delta\epsilon_{\text{BS}}$ and $\text{Im}\,\Delta\epsilon_{\text{BS}}$) in real space encoding high-$k_z$ phase and absorption, respectively. 
The spectral gap between two BS bands introduces oscillations in $z$-profiles. To obtain smooth BS $z$-profiles, a single BS band is isolated, and its magnitude, $|\Delta\epsilon_{\text{BS, sb}}|$, gives the oscillation envelope. Accordingly, a single BS band is used for $z$-profiles, whereas both bands are combined for phase-absorption decoupling or for $x$–$y$ cross-sections.
In contrast, the reflection-only geometry without the substrate accesses only a single BS band [Fig.~\ref{fig:SimulationRecon}(b)] and therefore cannot separate the phase and absorption contributions in the reconstruction.

Figure~\ref{fig:SimulationRecon}(e) and (f) shows the $x$–$z$ cross-sections of reconstructed $\Delta\epsilon_{\text{FS}}$ and $\Delta\epsilon_{\text{BS}}$ obtained using substrate-enhanced geometry. Since the reconstructed spectra obey the axial KK relation along $k_z$, both $\Delta\epsilon_{\mathrm{FS}}$ and $\Delta\epsilon_{\mathrm{BS}}$ are confined to the upper half-space ($z<0$); for clarity, the $z>0$ regions, which are nearly zero, are truncated in figures.

The reconstructed $\Delta\epsilon_{\mathrm{FS}}$ reveals smooth volumetric features similar to transmission, offering high lateral resolution but suffering from axial elongation and blurred boundaries due to the missing-cone in the low-frequency region [white shading in Fig.~\ref{fig:SimulationRecon}(d)], with real and imaginary parts corresponding to phase (left) and absorption (right), respectively.
The absence of high-$k_z$ components causes the top and bottom shell interfaces to vanish in $\Delta\epsilon_{\mathrm{FS}}$. 
In contrast, the recovered $\Delta\epsilon_{\mathrm{BS}}$ shows enhanced sensitivity to lateral interfaces, as shown in Fig.~\ref{fig:SimulationRecon}(f) and its inset, effectively recovering the missing boundaries of the shell and thereby complementing the FS information. 

Phase–absorption separation is evident in $\Delta\epsilon_{\mathrm{FS}}$ and $\Delta\epsilon_{\mathrm{BS}}$ under the substrate-enhanced geometry. In particular, for $\Delta\epsilon_{\mathrm{BS}}$, the real component predominantly captures phase interfaces, while the imaginary component captures absorptive interfaces. A continuous envelope combining both interfaces is shown in $|\Delta\epsilon_{\text{BS, sb}}|$.
For comparison, the results from the reflection-only geometry without the substrate are shown in Fig.~\ref{fig:SimulationRecon}(c), confirming that the BS features reconstructed and separated by our method are consistent with the reflection measurements.
However, only a single BS band is accessible in reflection-only geometry, analogous to optical coherence tomography (OCT).
This band limitation causes phase and absorptive interfaces to appear in both the real and imaginary components of the reconstructed $\Delta\epsilon_{\text{BS}}$, as shown in Fig.~\ref{fig:SimulationRecon}(c) and its inset, thereby precluding phase–absorption separation.

Results from the proposed method are further validated by comparison with a rigorous MBS-based reconstruction (Supplement 1). While yielding consistent results, the MBS approach incurs a substantially higher computational cost (>8 h), compared with the proposed framework, which requires approximately 100 s for inverse-matrix initialization and 5 s for reconstruction on an NVIDIA L40S GPU.

\subsection{Robustness analysis}

The robustness of the proposed reconstruction is numerically validated in Supplement 1 through systematic simulations assessing SNR, illumination pattern, temporal coherence, and multiple scattering, and is summarized here.

By adding white Gaussian noise to the scattering patterns to control the SNR, speckle-like artifacts gradually appear in the reconstructed $\Delta\epsilon_{\text{BS}}$. 
These speckle-like artifacts are analogous to those observed in OCT. As the added noise propagates through the linear inverse operator, the reconstructed fluctuations are filtered by the system transfer function, resulting in an autocorrelation that reflects the NA-dependent lateral and axial resolutions.
For variations in SNR and scanning density, the results show that FS–BS separation and phase–absorption decoupling remain stable down to an SNR of 23 dB and are largely insensitive to the specific illumination pattern, provided sufficient Fourier-space coverage is maintained. This behavior is consistent with the slow decay of the system’s singular values.

Analysis of different illumination strategies further reveals that scanning patterns produce distinct sampling densities in 3D Fourier space, particularly for the BS channel. Certain patterns introduce gaps in Fourier coverage, leading to reconstructions that preferentially emphasize specific structural features. Among the tested strategies, scanning schemes that approach a uniform distribution on a spherical cap, such as Fermat’s spiral, provide the most balanced overall performance~\cite{taddese2021optimizing}.

Temporal coherence effects are evaluated by synthesizing scattering patterns from four shell-shaped scatterers, which are more sensitive to BS, positioned at different depths and illuminated over a finite spectral bandwidth. The results indicate that, in the proposed method, the axial extent of the BS reconstruction is primarily limited by the temporal coherence length of the illumination source.

Finally, performance beyond the single-scattering regime is examined by increasing both scatterer density and permittivity contrast. Rather than failing abruptly, the method degrades gradually as multiple scattering strengthens, introducing increasing random phase variations from higher-order scattering paths that manifest as speckle noise and interface distortion. Nevertheless, interface information remains resolvable up to $\Delta\epsilon = 0.054$ at a volume fraction of 15\%.

\section{Experimental validation}

\begin{figure*}[t]
\hspace{-2cm}
\includegraphics[width=17cm]{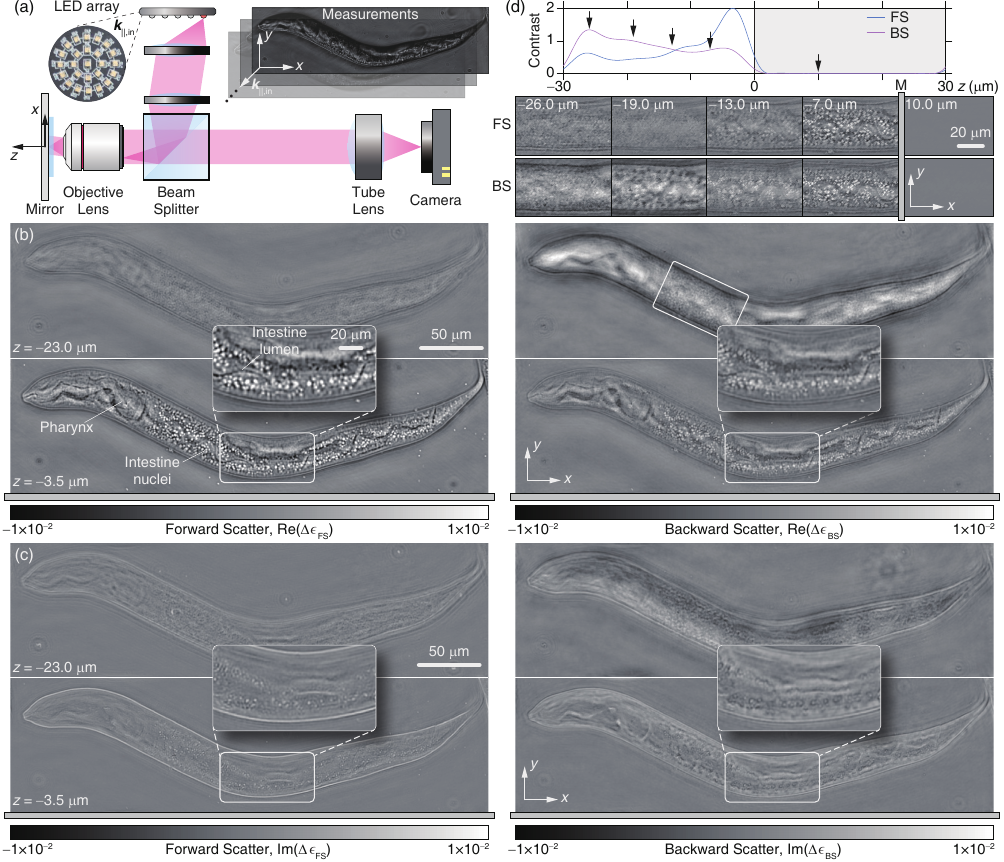}
\caption{\label{fig:Celegans}
Experimental demonstration with a fixed \textit{C.~elegans} on a mirror.
(a)~Schematic of the substrate-enhanced epi-illumination LED microscope; inset: LED array and representative measured scattering patterns.
(b) and (c)~Real and imaginary part of $x$–$y$ cross-sections of $\Delta\epsilon_{\mathrm{FS}}$ and $\Delta\epsilon_{\mathrm{BS}}$ at $z=-23.0~\mu$m (surface) and $z=-3.5~\mu$m (interior).
(d)~Depth-dependent feature contrast of $\Delta\epsilon_{\mathrm{FS}}$ and $\Delta\epsilon_{\mathrm{BS}}$, showing that the reconstruction is confined to the upper half-space. Inset, zoomed views at five depths.
}
\end{figure*}

\subsection{System setup and resolution characterization}

The technique was validated experimentally using a reflection-mode LED microscope with a 10$\times$/0.28NA objective  [Fig.~\ref{fig:Celegans}(a)]. Illumination was provided by a 25-LED array, relayed by a 4$f$ system to the objective’s back focal plane, generating plane waves with NAs of 0.14 and 0.28. Each LED emitted at 632 nm or 515 nm (20 nm bandwidth), and the resulting intensities were recorded by a camera (2.74~$\mu$m pixel size), as shown in the inset of Fig.~\ref{fig:Celegans}(a).

The setup was first validated using two-layer resolution targets and 3D randomly distributed beads (Supplement 1), demonstrating that FS and BS provide identical axial resolution and the same twofold lateral resolution enhancement relative to normal incidence, yielding an overall threefold increase in accessible bandwidth compared with the transmission geometry.

\subsection{Imaging of a moderately scattering biological specimen}

Next, experiments were performed on a fixed young \textit{C. elegans} worm, green alga \textit{Chlamydomonas}, breast cancer cells, and red blood cells, all placed on a silver mirror and immersed in water ($\epsilon_{\mathrm{m}} = 1.80$), with results for the latter two provided in Supplement~1.  
Among these samples, the young \textit{C. elegans} worm exhibits rich morphological features and operates slightly beyond the single-scattering regime, enabling validation of FS–BS imaging under moderate multiple scattering in a biological specimen. In contrast, the alga and red blood cells possess relatively simple structures and intrinsic absorption contrast, making them well suited for demonstrating phase–absorption separation.

The \textit{C. elegans} specimen was measured at a wavelength of 632 nm, and the transfer-function–based reconstruction was applied to recover $\Delta\epsilon_{\mathrm{FS}}$ and $\Delta\epsilon_{\mathrm{BS}}$. Representative $x$–$y$ cross-sections of the reconstructed phase (real part) and absorption (imaginary part) are shown in Fig.~\ref{fig:Celegans}(b) and (c), together with depth-dependent contrast profiles and a zoomed-in $z$-stack in Fig.~\ref{fig:Celegans}(d). Owing to the limited NA, the reported $z$ values represent approximate depths and do not provide accurate thickness information.

For the BS reconstructions, both BS bands were combined to generate the $x$–$y$ slices in Fig.~\ref{fig:Celegans}(c), whereas only the upper BS band was used to compute the contrast profile in Fig.~\ref{fig:Celegans}(d). This contrast profile, defined as $\sum_{\mathbf{r}_{\parallel}}|\Delta\epsilon|^{2}$, decays rapidly beyond $z=0$, confirming that the axial KK relation confines the reconstruction to the upper half-space.

Near the worm’s top surface ($z=-23.0~\mu$m), the reconstructed $\Delta\epsilon_{\mathrm{FS}}$ exhibits blurred, low-contrast interfaces due to the missing-cone effect, whereas $\Delta\epsilon_{\mathrm{BS}}$, arising from interfacial reflections, recovers high-frequency details and reveals pronounced, continuously varying surface features that complement the FS reconstruction. Deeper within the specimen ($z=-3.5~\mu$m), $\Delta\epsilon_{\mathrm{FS}}$ delineates internal anatomical structures, including the pharynx, intestinal lumen, and nuclei, while the BS contrast diminishes but remains spatially correlated with FS.

Comparison with bright-field images acquired in both transmission and reflection (Supplement 1) shows good agreement with the reconstructed FS and BS features, demonstrating that the proposed method effectively separates internal volumetric structures from surface-reflection–dominated features. Although reconstruction quality degrades in the presence of multiple scattering or finite temporal coherence length, these results indicate that the method remains effective for moderately scattering biological samples, providing high-throughput, complementary FS–BS information and a robust initial solution suitable for high-speed imaging or subsequent multiple-scattering–based reconstruction.

\subsection{Phase–absorption separation in biological samples}

\begin{figure}[t]
\centering
\includegraphics[width=7.5cm]{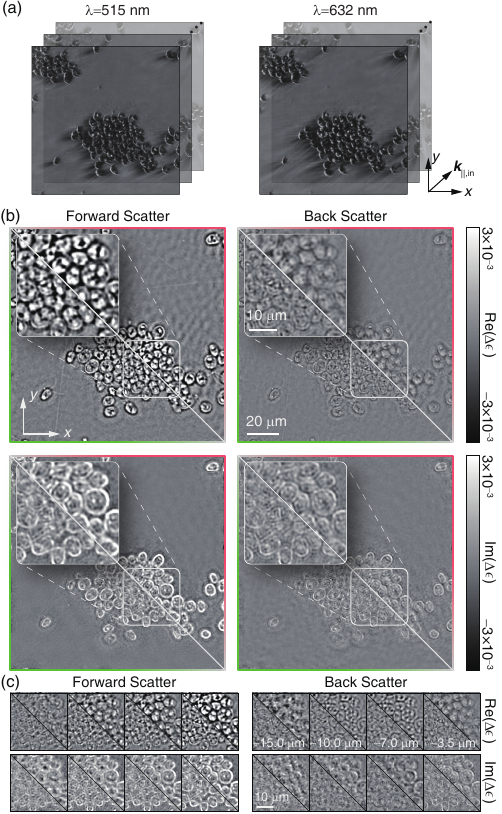}
\caption{\label{fig:GreenAlga}
Experimental demonstration with a cluster of \textit{Chlamydomonas} on a mirror.
(a)~Measured scattering patterns at 515 nm (left) and 632 nm (right).
(b)~Representative real and imaginary parts of $x$–$y$ cross-sections of $\Delta\epsilon$, with reconstruction at 515 nm in bottom-left panels and that at 632 nm in top-right panels.
Insets, zoomed views for detailed comparison.
(c)~$z$-stack profiles of the framed regions in (b).
}
\end{figure}

To experimentally demonstrate phase–absorption separation, a cluster of \textit{Chlamydomonas}, which appears visibly dark green under sunlight, was placed on a mirror substrate and sequentially imaged at wavelengths of 515 nm (green) and 632 nm (red), as shown in Fig.~\ref{fig:GreenAlga}(a).

Representative $x$–$y$ cross-sections at $z=-2.2~\mu$m reconstructed at the two wavelengths are presented in Fig.~\ref{fig:GreenAlga}(b), with the 515 nm reconstructions shown in the bottom-left panels and the 632 nm reconstructions in the top-right panels. FS reconstructions derived from the FS band are shown in the left column, while BS reconstructions combining both BS bands are shown in the right column. Framed regions are enlarged as insets for detailed comparison, and their corresponding $z$-stack profiles are shown in Fig.~\ref{fig:GreenAlga}(c).

The real part of the reconstructed $\Delta\epsilon$ exhibits consistent phase contrast at both wavelengths, clearly resolving the spherical cell body and flagella. These phase features encode internal cellular morphology in FS and interfacial information in BS. In contrast, the imaginary part of $\Delta\epsilon$ reveals pronounced absorption features that predominantly surround internal structures. The absorption contrast is distinct from the phase distribution and is consistent with incoherent-illumination bright-field images shown in Supplement~1.
At 632 nm, both FS and BS reconstructions exhibit stronger absorption contrast, with mean values of $\overline{\mathrm{Im}\,\Delta\epsilon}=5.2\times10^{-4}$ and $2.4\times10^{-4}$ in the framed region, respectively. By comparison, reconstructions at 515 nm yield reduced absorption contrast, with corresponding values of $4.3\times10^{-4}$ and $1.9\times10^{-4}$. This wavelength dependence is consistent with the dark green appearance of \textit{Chlamydomonas}, which reflects green light while exhibiting stronger absorption at longer visible wavelengths associated with chlorophyll.

The same experiments were performed on red blood cells, with results shown in Supplement~1. In contrast to \textit{Chlamydomonas}, the reconstructed absorption contrast of red blood cells is higher at 515 nm than at 632 nm, consistent with their red appearance and the stronger absorption of green light by hemoglobin.
Together, these results experimentally validate robust phase–absorption separation in both FS and BS channels using the proposed technique.

\section{Conclusion and discussion}
In conclusion, we have introduced a substrate-enhanced diffraction tomography method that simultaneously retrieves FS and BS in epi-configuration, mitigating the missing cone of transmission and tripling the accessible 3D Fourier bandwidth. FS resolves internal volumetric structures, while BS captures high-frequency surface interfaces, providing complementary contrast confirmed in simulations and experiments. We derive explicit 3D transfer functions for FS and BS and establish an axial KK relation that constrains inversion to the physical half-space. These transfer functions enable band-limited reconstruction restricted to the physical passbands, substantially reducing computational cost. The approach achieves high-resolution, label-free volumetric imaging, disentangles FS–BS mixing in epi-geometry, and extends the effective bandwidth of conventional configurations.

In future work, we anticipate several extensions of the proposed framework.  
First, for objectives with moderate NA, the substantial spectral separation between the FS and BS bands allows them to be analyzed independently. As the NA increases, this separation progressively diminishes, until the axial bandwidth satisfies $B_z \ge 4k_m/3$ (corresponding to an illumination angle of approximately $71^\circ$). At this point, the central FS band and the two BS bands merge to form a 4Pi bandwidth~\cite{mudry2010mirror}, theoretically enabling isotropic resolution in all spatial directions, as demonstrated in Supplement~1. This suggests that the proposed approach may provide a practical route to achieving 4Pi tomographic bandwidth using a single objective lens and an interference-free optical path.

Second, the proposed technique exhibits a dependence on NA and temporal coherence length that is complementary to that of spectral-domain (SD) OCT~\cite{zhou2021unified}. In our framework, increasing the NA expands the axial BS spatial-frequency support and improves axial resolution, while a relatively narrow spectral bandwidth enables an extended axial reconstruction range. Consequently, limitations imposed by finite temporal coherence length can be mitigated by employing laser illumination in future. In contrast, SD-OCT relies on a broad bandwidth for high axial resolution, while favoring a low NA to maintain an extended axial imaging range. These complementary dependencies suggest that integrating multi-spectral illumination with angle-diverse measurements could further enhance the recovery of BS information~\cite{zhang2018multi}. 

Finally, the substrate, present in nearly all reflection-mode imaging systems, can be treated as an additional controllable degree of freedom for engineering the scattering field. By tailoring its optical properties, the substrate may selectively suppress specific scattering components, enhance BS sensitivity, or generate passive structured illumination. Such substrate-assisted control offers a flexible strategy to optimize axial frequency support and scattering contrast without modifying the illumination or detection geometry, opening new avenues for high-resolution volumetric reconstruction in reflection-mode imaging.

\begin{backmatter}
\bmsection{Funding}
The work was supported by National Science Foundation (1846784) and in part by grant number 2023-321163 from the Chan Zuckerberg Initiative DAF, an advised fund of Silicon Valley Community Foundation.

\bmsection{Acknowledgment}
The authors thank Boston University Shared Computing Cluster for proving the computational resources. 

\bmsection{Disclosures}
The authors declare no conflicts of interest.

\bmsection{Data Availability Statement}
The reconstruction algorithm is available at: \href{https://github.com/bu-cisl/TF\_rIDT}{https://github.com/bu-cisl/TF\_rIDT}.

\bmsection{Supplemental document}
See Supplement 1 for supporting content.

\end{backmatter}

\bibliography{sample}

\end{document}